# Tracking atomic structure evolution during directed electron beam induced Si-atom motion in graphene via deep machine learning

Maxim Ziatdinov[1,2], Stephen Jesse[1], Bobby G. Sumpter[1], Sergei V. Kalinin[1], Ondrej Dyck[1]

[1]*Center for Nanophase Materials Sciences,*
*Oak Ridge National Laboratory, Oak Ridge TN 37831*
[2]*Computational Sciences and Engineering Division,*
*Oak Ridge National Laboratory, Oak Ridge TN 37831*

**Abstract:**

Using electron beam manipulation, we enable deterministic motion of individual Si atoms in graphene along predefined trajectories. Structural evolution during the dopant motion was explored, providing information on changes of the Si atom neighborhood during atomic motion and providing statistical information of possible defect configurations. The combination of a Gaussian mixture model and principal component analysis applied to the deep learning-processed experimental data allowed disentangling of the atomic distortions for two different graphene sublattices. This approach demonstrates the potential of e-beam manipulation to create defect libraries of multiple realizations of the same defect and explore the potential of symmetry breaking physics. The rapid image analytics enabled via a deep learning network further empowers instrumentation for e-beam controlled atom-by-atom fabrication. The analysis described in the paper can be reproduced via an interactive Jupyter notebook at https://git.io/JJ3Bx

Scanning transmission electron microscopy (STEM) has become one of the most powerful tools for exploring materials structure on the atomic scale. Image data collected with STEM contains encoded information about configurations and interactions of single atomic defects,[1-4] structure of domain walls and interfaces,[5-8] and internal electric fields on the atomic level.[9] However, most of the STEM studies analyze atomic configurations observed in a single image, leaving aside the possibility for minor variations which may manifest in a more broadly applied statistical analysis of many acquired images of distinct examples of the same structural configuration. Furthermore, there can be hidden degrees of freedom – "impurities" that cannot be revealed directly from static images (unless extensive theoretical modelling is performed) but that may result in additional distortions of certain atomic structures – such as in the case of "invisible" OH groups in the Pt/γ-alumina catalytic system[10].

Recently, STEM was shown to be a powerful tool for moving atoms. Following predictions that STEM might be utilized to produce controlled atomic motion,[11, 12] a series of results have demonstrated this is true. In graphene, Si dopant atoms have been controllably inserted into the lattice,[13] moved through the lattice,[13-15] moved along graphene edges and incorporated into the lattice by attaching to edges and subsequently growing the graphene lattice *in situ*.[16] These investigations have culminated in the recent demonstration of atom-by-atom assembly of primitive structures embedded in graphene.[17] This level of control has yet to be extended to other 2D materials, however exciting results have recently been published where similar controlled atomic motion was achieved in a bulk Si crystal.[18, 19] While demonstrating atomic plane precision crystallization of Si, the authors showed that Bi dopants grown into the crystal could be moved into a line by using a variant of the crystallization proceedure. This phenomenon was investigated in greater detail and the ability to position the Bi dopants with atomic column precision was demonstrated as well as formation of Bi clusters and patterns.[18] These examples illustrate the remarkably precise alterations accessible to STEM-based manipulation modalities. However, additional investigations are necessary to unravel the precise configurations and subtle alterations involved in e-beam modification of materials.

Here, we combine direct atomic e-beam manipulation with deep machine learning based analysis to extract material evolution at the atomic level during manipulations. Specifically, we realize long-range linear and rotational motion through a graphene lattice. This allows us to collect multiple statistically independent configurations for individual defects, track changes in the specific atomic bonds in time, and determine whether symmetry breaking is present in the system.

**Moving an impurity with an electron beam.**

A focused STEM electron beam can be used to move dopant atoms through a crystal lattice, as already mentioned. To accomplish this, the beam is positioned at a lattice site adjacent to the dopant. In the case of moving Bi in a Si crystal, the beam creates a vacancy in the adjacent column and induces a controlled diffusion of the Bi toward the beam.[18] In the case of Si in graphene, of interest here, the beam induces a bond rotation between the adjacent C and Si atom again resulting in the dopant moving toward the beam position.[12] Here, we use this method to move Si atoms through a graphene lattice and examine the lattice structure and relative atomic positions at each step.

Two experiments were performed where Si dopant atoms were moved through a graphene lattice using the electron beam and an image was acquired after each successive movement of the atom. These image sequences were concatenated to form videos of the atomic motion and are available via the accompanying notebook. Figure 1 shows a summary of the experimental data obtained. For the first experiment, the Si atom was moved repeatedly around a hexagonal ring in the lattice. Images were acquired as quickly as possible (while maintaining intensity) to decrease the likelihood of unintentional movement of the dopant. The image shown in a) is the result of averaging twelve well-aligned video frames together. The bright Si atom has traversed the hexagon twice, resulting in its increased intensity on average. This also allows a clearer view of each lattice site. Figure 1 b)-g) illustrate the motion of the Si dopant once around the hexagon and are taken from sequential frames in the associated video. The dot marks the electron beam position used to achieve the movement from one frame to the next. The arrow marks the Si position through time.

For the second experiment, shown in h)-i), a Si dopant was moved linearly from the lower left of the field of view to the upper right. h) shows the initial configuration and i) shows the final configuration with the Si position from each intermediate frame marked with a dot. In both these experiments the Si remained in the 3-fold coordination throughout, resulting in many images of 3-fold coordinated Si dopants but each image representing a different atomic structural configuration, as distinct from simply acquiring multiple images a defect. This gives us the opportunity to examine the structure for slight deviations from symmetry which may be uncovered from a statistical analysis of examples of the same defect.[20]

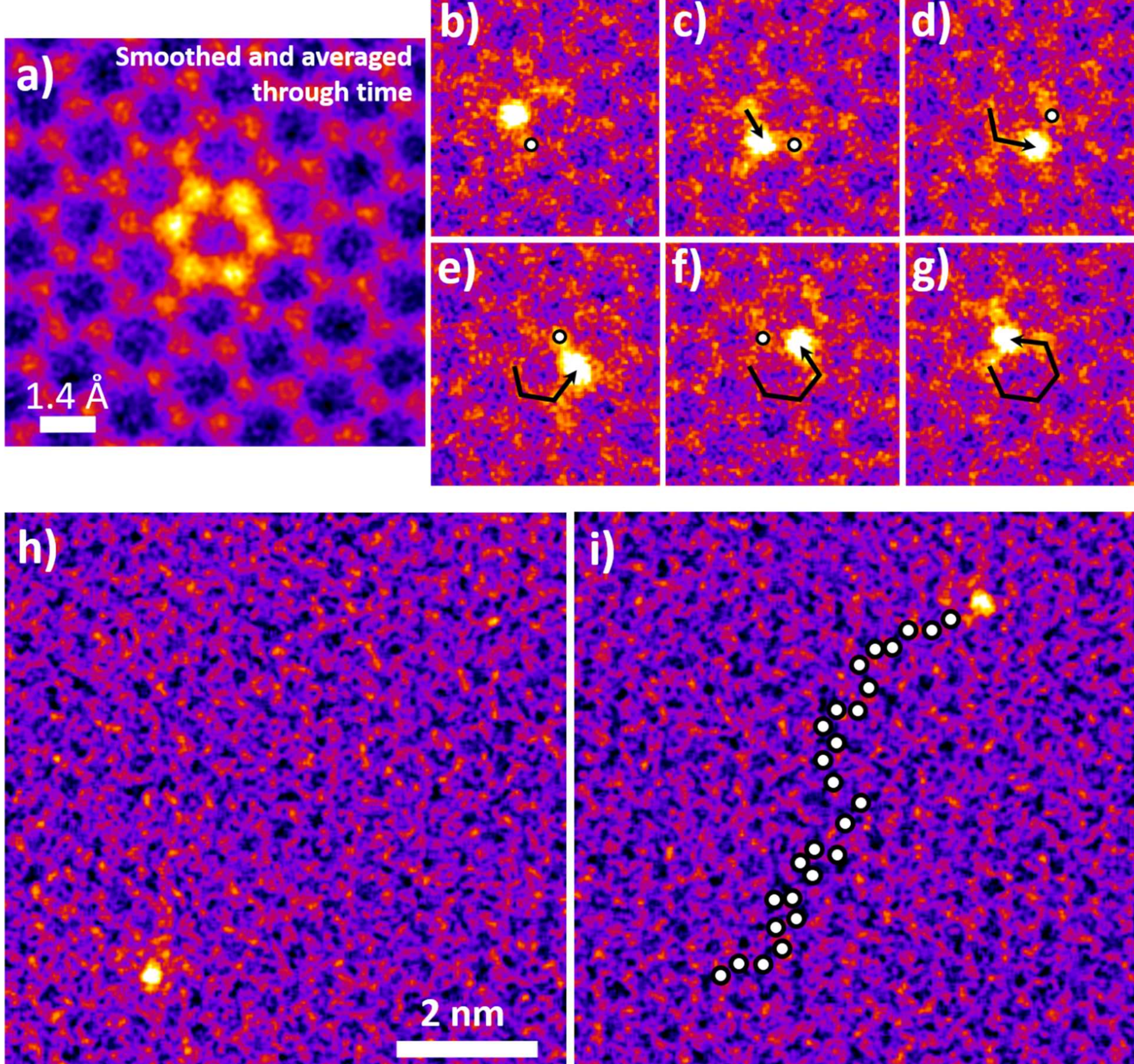

**Figure 1 Directed atomic motion of Si atoms through a graphene lattice using electron beam exposure.** The upper panel, a)-g), illustrates moving the Si atom around a graphene hexagon in circles. a) shows a set of images which have been smoothed (gaussian blur) and averaged through time to show how the Si atom, which appears bright, has occupied each position in the central hexagon. Images b)-g) illustrate the movement of the Si atom frame-by-frame. The dot marks the electron beam position used to induce the movement achieved in the following frame. The arrow marks the Si atom positions through time. The lower panel, h)-i), show movement of a Si dopant

linearly through the lattice. h) shows the initial configuration and i) shows the final configuration with each intermediate position marked with a dot. Noise was reduced using a gaussian blur. All images were artificially colored using the "Fire" look up table in ImageJ.[21]

We start by reconstructing ($x$, $y$)-positions of lattice atoms and impurity atoms for each frame of the first STEM movie. To achieve this, we trained a deep fully convolutional neural network (Figure 2) for locating atomic positions in noisy experimental data where atoms do not always appear as local maxima. The model takes a "raw" experimental 2D image/frame as an input and outputs the "probability" of each pixel in that image belonging to an atom (or a certain type of atom) or to the background. Our neural network has an encoder-decoder type of architecture with a bottleneck layer. The encoder part consists of multiple blocks of convolutional layers followed by a max-pooling operation. The decoder part has the same convolutional blocks but in the reversed order and up-sampling operation via the bilinear interpolation before each of them. The bottleneck layer represents a spatial pyramid of dilated convolutions with dilated rates 2, 4, and 6. The network classifies every pixel in the raw data and outputs a set of well-defined circular blobs on a uniform background whose centers of mass correspond to the atomic centers. This enables a bijective (i.e., one-to-one) mapping of the extracted features with the original input data. The ($x$, $y$) positions of atoms can be (optionally) further refined by using the blob centers of mass from the network's output as the initial guess and performing a standard 2D Gaussian peak fitting within the specified area (e.g., within a box with a side equal to ½ of an average nearest neighbor distance) on the raw input image pixels. We note however that this approach may not improve the accuracy of atom finding for noisy data where there are no clear maxima associated with atomic positions. The number of classes in the output can be equal to the number of different atomic species that one expects to observe in the experiment. However, it is usually difficult predict beforehand which chemical species will be observed (e.g., due to the presence of contaminations). The alternative strategy, which was used in this work, is first to categorize all atoms as a single class and then perform an additional post-processing step where atoms are separated into different classes based on the statistical analysis of their intensities or of their local neighborhood.

The neural network was trained to recognize atoms in graphene atomic structures based on a library of MultiSlice[22] STEM image simulations of a graphene lattice with different configurations and impurities.[23] The simulated data was further augmented to account for instrumental factors such as variations in the level of noise, drift, etc. A total number of images used for training was ~3000. The model weights were trained using Adam optimizer with cross-entropy loss function. To prevent our model from overfitting, the dropout layers[24] were used in the deep layers. Interestingly, we found that when training on simulated data and applying to real data, the use of batch normalization[25] for regularization usually leads to a degraded performance and that the dropout layers should be used instead. The experimental images (video frames) were fed into a trained network without resizing to prevent the introduction of artificial distortions in the analysis. All the data analysis was performed using a home-built AtomAI software.[26]

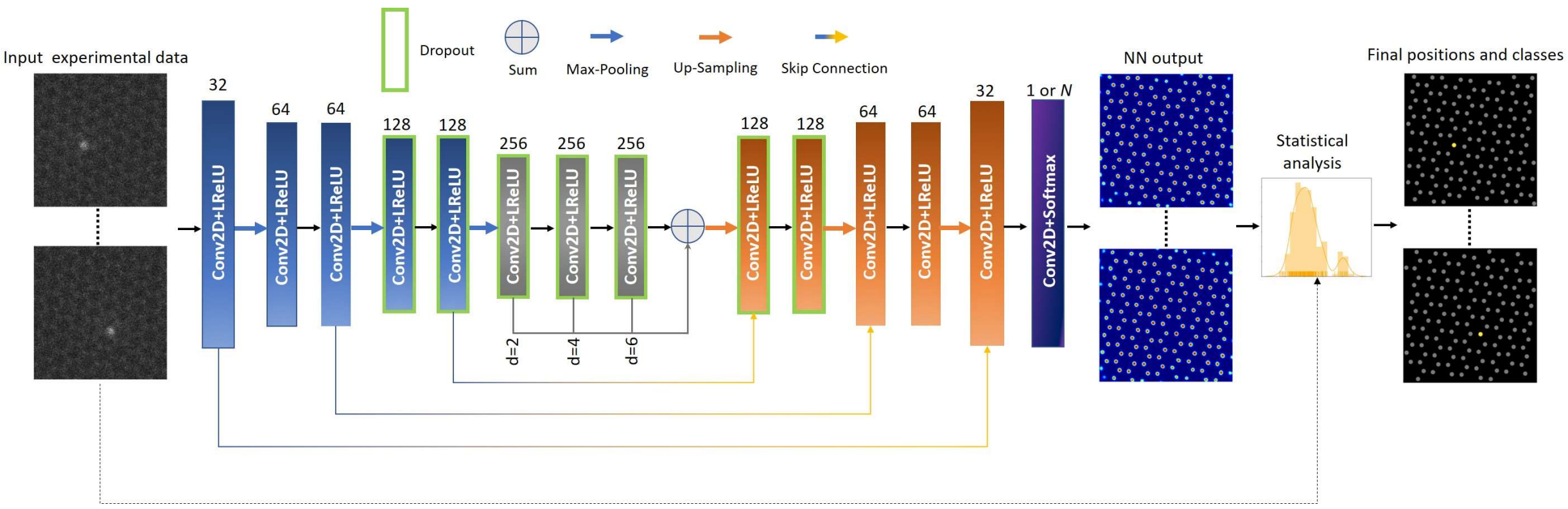

**Figure 2. Schematics of deep learning-based identification of atomic species in experimental data.** The fully convolutional neural network takes raw experimental data as an input and outputs pixel-wise classification maps of the “probability” of each pixel belonging to an atom. The number at the top of each block denotes the width of each layer (i.e., number of convolutional filters). For the final layer, this number corresponds to the number of classes. If the number of classes is 1, the network doesn’t distinguish between the atoms (different atomic species are categorized as one class). Alternatively, one can adopt a classification scheme with multiple $N$ atomic classes (in this case, pixels associated with *e.g.* C and Si atoms will be in different output “channels”). The statistical analysis can be performed on raw pixels using the atomic positions extracted from neural network output to update or refine the identified classes. The d=2, 4, 6 denote the dilation rate in each layer in the bottleneck block. The coloring scheme serves as categorical distinction for the eye.

We now start analysis of the experimental data by studying overall distribution of the atomic bonds as well as their spatial variation for each frame. To achieve this, we first mapped the output of the deep learning model onto a lattice graph and performed an automated search of nearest neighbors for each identified atom. This allowed us to calculate all the relevant atomic bond lengths. The histogram of bond lengths for all the frames of the movie is shown in Fig. 3a.

The bond lengths are normally distributed with a mean of 144.7 pm and a standard deviation of 10.4 pm. The mean value is very close to the graphene equilibrium lattice constant 142 pm. Next, we attempted to trace whether there is any specific bond distortion (local strain) that moves with a dopant atom. For this, we use the information extracted for atomic bond lengths to construct the real-space strain maps for one of movies. Here, we define strain as $s = (a-a_0)/a_0$, where $a_0$ is the mean value determined from data shown in Fig. 3a. The movies of strain evolution through time are available via the accompanying Jupyter notebook and the selected frames are shown in 3b. We found that generally variation in bond lengths with the field of view does not have a clear correlation with position of Si atom. This indicates that we were able to move the Si impurity without causing any "side effects" to the graphene lattice. Finally, we note that it takes only about 1 second to get from raw experimental data to bond maps (see the accompanying interactive notebook) meaning that information about minute atomic displacements in the lattice can be obtained on-the-fly.

We next analyze local atomic neighborhoods via Gaussian mixture model (GMM)[27] and principal component analysis (PCA)[28] (Figure 4a, b). Here, a stack of sub-images centered at carbon (or Si) atoms was clustered using the GMM method, revealing the presence of two well-defined classes corresponding to the two lattice sites of primitive graphene lattice. In this case, the unsupervised GMM analysis and prior knowledge of physics of the system unambiguously dictate the choice of atom classes. Within each class, the sub-images can be analyzed the PCA method, as explored earlier for the local crystallography[29] and FerroNet[30] approaches. The corresponding PCA components are shown in Fig. 4 and exhibit orthogonal patterns of atomic displacements. The analysis of spatial localization and temporal evolution of these patterns does not reveal any specific structure, as can be expected given the extremely slow (compared to the atomic dynamic) timescales of the STEM measurements. Hence in certain sense the PCA components in Fig. 4 represents the possible orthogonal functions in which the atomic patterns can be sampled via Gaussian noise addition. Finally, the ability to obtain and compare atomic coordinates of a Si-C complex from the many repeated observations of positioning the same atom in multiple nominally equivalent lattice sites can give an insight into the statistical details of this defect structure. We therefore performed a combined GMM + PCA analysis for the Si-C cluster specifically, which we define here as the Si impurity atom with the associated lattice atoms in its first two coordination spheres (Figure 4c, d). Here, the distortions associated with PCA components 3 and 4 in both sublattices may originate from a random symmetry breaking in Si-C bonds.

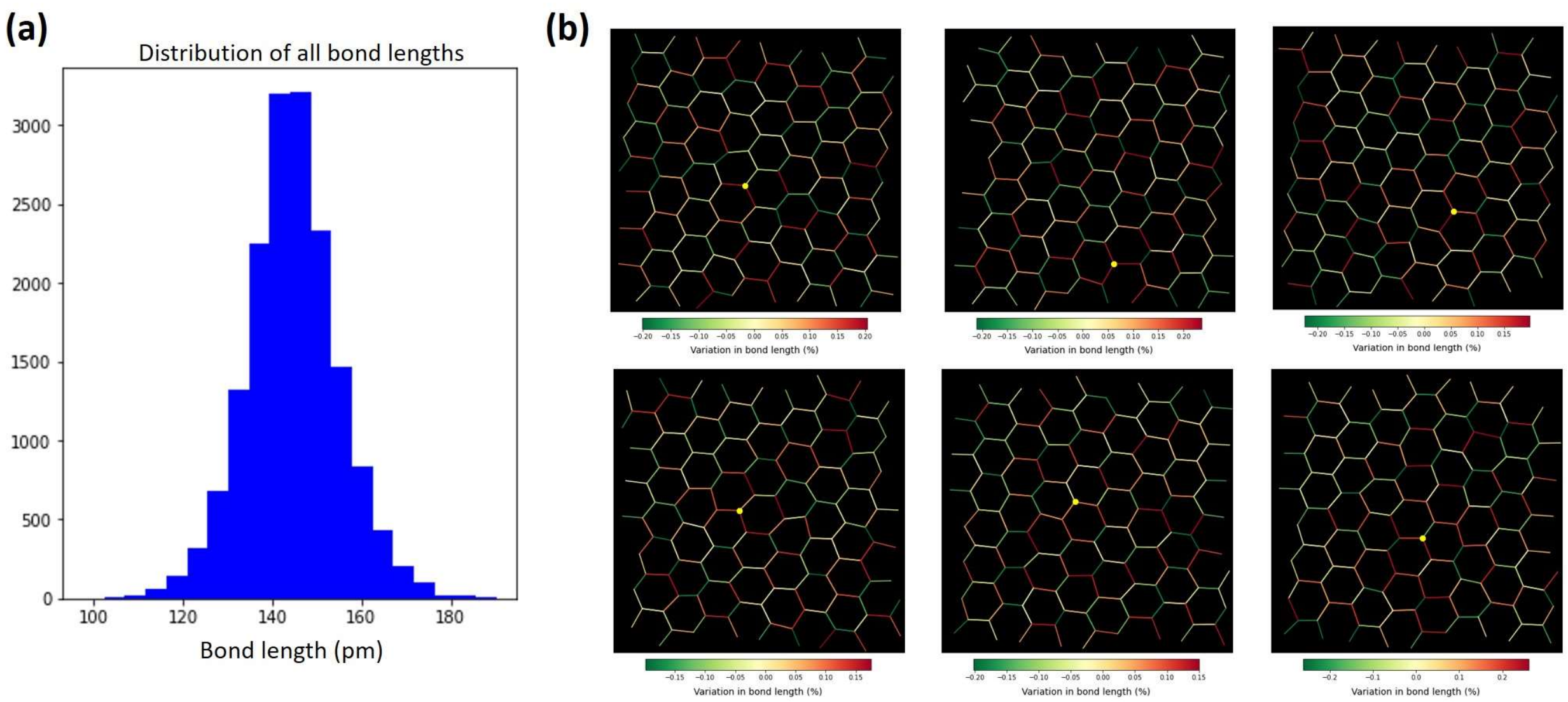

**Figure 3. Real-space mapping of atomic distortions in STEM movie data.** (a) Distribution of bond lengths for all the atoms in the first movie. (b) Strain maps for selected frames (top row: frames 5, 10, 15; bottom row: frames 35, 45, 50) of the first movie. The yellow dot denotes a position of the Si atom.

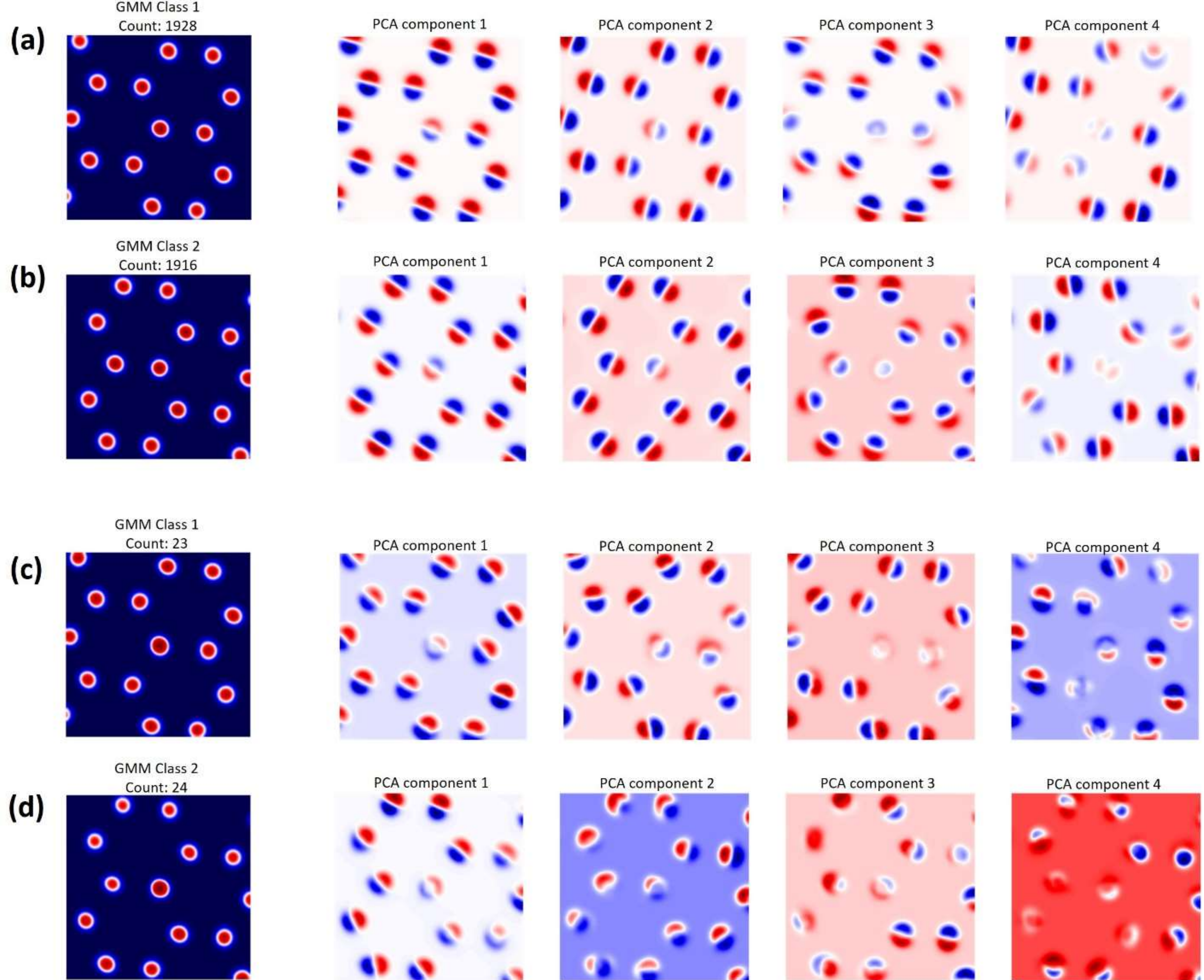

**Figure 4. Combined Gaussian mixture model (GMM) and principal component analysis (PCA) of local atomic neighborhoods.** (a, b). Analysis of local distortions in carbon lattice. The GMM allows two graphene sublattices to be separated into two different classes in the automated fashion. PCA can be then performed individually on each class to view possible characteristic atomic distortions. (c, d) The same analysis for the local neighborhood of Si impurity (a larger blob in the center of GMM components).

To summarize, we enabled the deterministic electron beam motion of individual Si atoms along predefined trajectories including a circular and linear trajectory. The structural evolution during the dopant motion was explored, providing information on changes of the Si atom neighborhood during atomic motion and providing statistical information of possible defect configurations. Overall, this approach demonstrates the potential of e-beam manipulation to create defect libraries of multiple realizations of the same defect and explore symmetry breaking physics.

The rapid image analytics enabled through the deep learning network further provides enabling instrumentation for e-beam atom-by-atom fabrication.

**Acknowledgments:**

This research was sponsored by the Division of Materials Sciences and Engineering, Office of Science, Basic Energy Sciences, US Department of Energy (RVK and SVK). Research was conducted at the Center for Nanophase Materials Sciences, which is a US Department of Energy Office of Science User Facility. STEM experiments (OD, SJ) and data analysis (MZ) were supported by the Laboratory Directed Research and Development Program of Oak Ridge National Laboratory, managed by UT-Battelle, LLC, for the U.S. Department of Energy.